\documentclass[11pt]{article}
\usepackage{graphicx}
\usepackage[left=1in, right=1in, bottom=1.2in, top=1.2in]{geometry}
\usepackage{graphicx}
\usepackage{caption,subcaption}
\usepackage{charter}
\usepackage{amsmath}
\usepackage{xcolor}
\usepackage[colorlinks=true, allcolors=blue]{hyperref}
\usepackage{siunitx}
\usepackage{lipsum}
\title{Interference with (Pseudo) Thermal Light; The Hanbury Brown and Twiss Effect} 
\date{}
\author{Km Nitu Rai,$^{1}{}^{*}$ Soumen Basak,$^{1}$ Subrata Sarangi,$^{2, 3}$ and	Prasenjit Saha$^{4}$}
\begin{document}
\maketitle
\noindent
\small{${}^{\textbf{1}}$School of Physics, Indian Institute of Science Education and Research Thiruvananthapuram, Maruthamala \\PO, Vithura, Thiruvananthapuram 695551, Kerala, India\\
$^\textbf{{2}}$School of Applied Sciences, Centurion University of Technology and Management, Odisha-752050, India \\
$^\textbf{{3}}$Visiting Associate, Inter-University Centre for Astronomy and Astrophyics, Ganeshkhind, Savitribai Phule Pune University Campus, Pune 411 007, India \\
$^\textbf{{4}}$Physik-Institut, University of Zurich, Winterthurerstrasse 190, 8057 Zurich, Switzerland}

\let\thefootnote\relax
\footnotetext{$^{*}$\textmd{niturai20129617@iisertvm.ac.in}} 

\begin{abstract}
The correlation of light from two sources leads to an interference pattern if they belong to a specific time interval known as the coherence time, denoted as $\Delta \tau$. The relationship governing this phenomenon is $\Delta \tau \Delta \nu \approx 1$, where $\Delta \nu$ represents the bandwidth of the light. This requirement is not satisfied, and hence, interference fringes are not observable in the case of ordinary (thermal) light. In the 1950s, Robert Hanbury Brown and Richard Q. Twiss explored interference phenomena using a narrow bandwidth of thermal light. This investigation led to the discovery of the Hanbury-Brown and Twiss effect (or the HBT effect in short), which has since found applications in various fields, particularly stellar observations and quantum optics. This article briefly traces the history of the HBT effect and its applications in various fields, including stellar observations. More importantly, it outlines the basic theoretical framework of this effect, followed by the design and results of the correlation in intensity fluctuation of a pseudo-thermal light in a college laboratory setting (Michelson interferometer).\\
\noindent
\textbf{Keywords:} Michelson Interferometer, pseudo-thermal light, auto-correlation
\end{abstract} 

\bigskip
\section{Introduction}\label{sec:intro}
The most common phenomenon in optics is interference, where two coherent beams overlap and produce a static intensity distribution called interference fringes. Different setups exist to study the optics using this phenomenon. One of the popular setups is the ``Michelson interferometer", invented by American physicists Albert A. Michelson and 
Edward W. Morley and can be built easily with a monochromatic source, lens, beam-splitter, and two reflecting mirrors. This precision instrument produces an interference pattern by splitting a light beam with the beam-splitter into two parts and recombining them after making them travel through different optical paths. Historically, the Michelson Interferometer was used to check the hypothesis of the luminiferous aether medium, a supposed medium permeating the space that was thought to be the carrier of light waves. 

There are many variants of the Michelson interferometer in astronomy. 
The most recent is the ``Laser Interferometer Gravitational-Wave Observatory (LIGO)", which uses gravitational waves to split in two beams with the beam-splitter and demonstrates distortion in space-time caused by mergers of massive neutron stars and black holes \cite{abbott2016gw151226}. Another type is the stellar interferometer of Michelson and F. G. Pease, which is referred to as the "Michelson Stellar Interferometer". This interferometer employed a pair of plane mirrors separated by a 20-feet-long baseline with the 100-inch reflector at Mount Wilson and measured the angular diameter of Betelgeuse ($\alpha$-Orionis) \cite{michelson1921measurement}. This interferometer did not use any beam splitter; two separate mirrors were used to collect the photons arriving from the star and were then made to interfere. The result of this instrument set the birth of a new branch in Observational Astronomy called Amplitude Interferometry. This technique has a wide variety of applications in Astronomy. In fact, several telescopes in Astronomy nowadays use this technology to operate at different wavelengths of light. However, we will not discuss these variants in the present article.

We will discuss an alternative that dispenses with the requirement of coherent light. Exploring this procedure and following the analysis of the radio galaxy Cygnus and a supernova remnant Cassiopeia using radio waves in the 1950s \cite{hanbury1952apparent}, R. Hanbury Brown and R. Q. Twiss proposed the novel idea of measuring the correlation between the intensity fluctuations of two receivers of the same type of instrument operating in radio frequencies \cite{brown1954lxxiv}. This procedure was then named the Hanbury Brown and Twiss (HBT) effect \cite{brown1957interferometry}, honoring their work.

According to this effect, the second-order correlations, i.e., correlations among intensities fluctuations, carry the source information. Based on this effect, the Intensity Interferometry (II) is considered to be an Interferometer with an incoherent light source. In II, instead of measuring the electric field amplitudes of the wave disturbances received on the detector, the intensity is measured within a small bandwidth $\Delta \lambda $.\footnote{Before the emergence of the HBT effect, the light was traditionally viewed as a collection of individual photons, and Hanbury Brown detected correlations in photons from two detectors. It created a famous controversy in the science of photons. There were many theoretical and experimental proofs in support and opposition to the HBT effect. One of them, German-American physicist Leonard Mandel, presented his theoretical findings regarding the HBT effect, namely that the number of photons arriving within a specific time interval adhered to the pure Bose-Einstein distribution. The controversy was over, and the new field of Optics, Quantum Optics, was born.} More precisely, photons received at two detectors within a time $\Delta \tau = \frac {1}{\Delta \nu} = \frac{\lambda ^2}{c \Delta \lambda}$ correlate (or anticorrelate) with each other. Hanbury Brown and Twiss also reported the results of their laboratory experiment \cite{brown1956correlation} designed to validate the theory of correlation in intensity fluctuations. This experiment used a combination of a mercury arc, a liquid filter, and a pinhole as the source of light, two photomultipliers coupled with amplifiers to detect and amplify the photon signals, and a correlator to compute the intensity (or photon) correlation. Subsequently, they experimented with partially coherent light to verify the theory of measuring astronomical sources \cite{hanbury1979test, brown1958interferometry}. In the 1960s and 1970s, their idea was successfully implemented in particle physics, ultimately leading to a new era of quantum optics \cite{glauber1963quantum}. 

Later, in 1974, the historic Narrabri Stellar Intensity Interferometry (NSII) was built based on the HBT effect. NSII observed 32 stars and measured their diameters \cite{hanbury1974angular}. The primary goal of building this instrument was to deal with atmospheric turbulence and to work on longer baselines. Since then, research in the area of Quantum Optics has progressed steadily; however, there came a lull in Stellar Intensity Interferometry due to the limitations in the sensitivities of the then-available photon detectors. Thankfully, with more sensitive and faster photon counters available over recent years, Intensity Interferometry (II) is getting back into use \cite{davis1999sydney}. The rebirth of solar neighborhood exploration using II is being reported in astronomy literature.

This Resonance article discusses the HBT effect that emerges from the second-order correlation of the intensity distribution. The intensity distribution is produced by a Michelson Interferometer setup using a pseudo-thermal light beam following correlation in amplitude. The numerical analysis of intensity distribution over time returns the result as the HBT effect.
\section{Theoretical Background}\label{sec:theory}
\begin{figure}[hbt]
	\includegraphics[width=\linewidth]{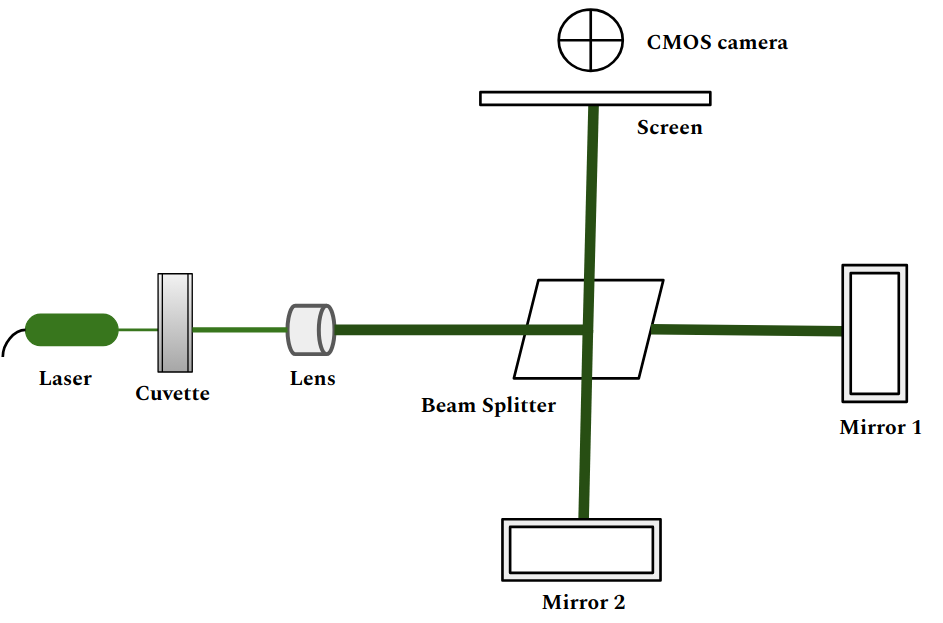}
	\caption{Schematic of the Michelson interferometric setup used in our experiment is shown. The cuvette is filled with distilled water and a few drops of milk. The laser beam emerging from the cuvette emulates thermal light. The CMOS camera behind the screen captures the intensity distribution over time.}
	\label{fig:MI}
\end{figure}
\begin{figure}[hbt]
	\includegraphics[width=\textwidth,height=4in]{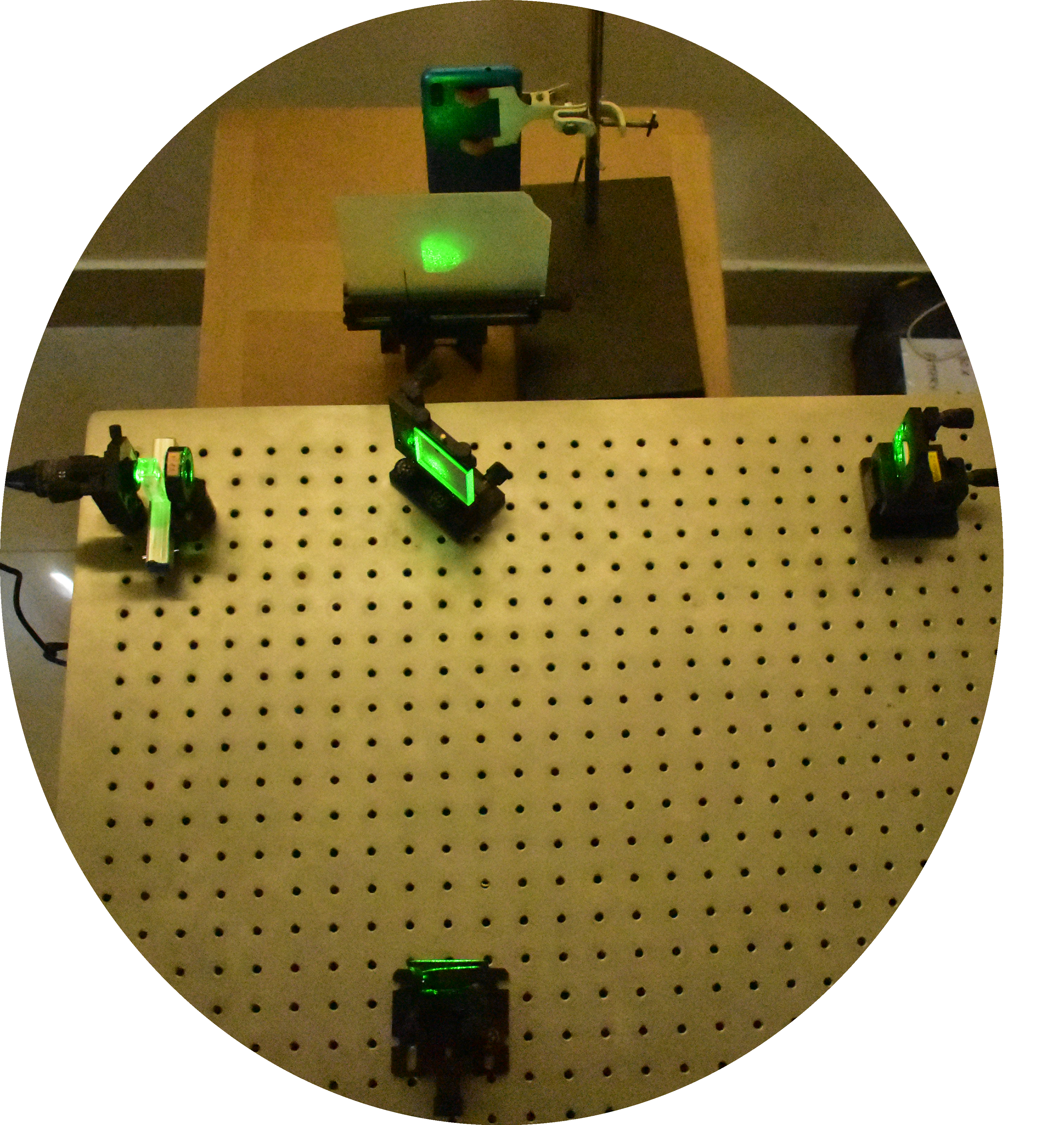}
	\caption{A photograph of our Michelson Interferometer set up with pseudo-thermal light in the laboratory. The laser beam-cuvette-lens combination, which produces the pseudo-thermal light, is seen on the left, almost forming a single unit. The CMOS camera of the smartphone placed behind the screen is used to generate video records of the intensity pattern visible on the screen.}
	\label{fig:MII}
\end{figure}
In this section, the derivation of intensity distribution for coherent and incoherent light will be studied using a Michelson interferometer setup.  

\subsection{Working with coherent light:}

Let us take a coherent laser light with wavelength $\lambda$ and angular wavenumber $k=2\pi/\lambda$. This beam of light gets split by a beam splitter (See Fig.~\ref{fig:MI}). The two secondary beams, thus created, get reflected at the two mirrors, travel back to the beam splitter through two different optical paths (as designed), pass through the beam splitter, and recombine on the screen.

On the screen at a point, say $x$, the wave amplitude due to the light beam reflected from Mirror~1 will have the form
\begin{equation}
    a_1 e^{i\phi_1 - ikl_1(x)}
    \label{eqn:single}
\end{equation}
where $l_1(x)$ is the path length for that light ray, while $a_1e^{i\phi_1}$ represents the rest of its complex amplitude.  An analogous expression applies to the light reflected from Mirror~2.  Adding the two contributions, the complex amplitude at point $x$ on the screen will be
\begin{equation}
A(x) = a_1 e^{i\phi_1 - ikl_1(x)} + a_2 e^{i\phi_2 - ikl_2(x)}
\label{eqn:ampl}
\end{equation}
The resulting intensity at $x$ will be
\begin{equation}
I(x) = a_1^2 + a_2^2 + 2a_1a_2 \cos[k\,l_1(x)-k\,l_2(x)+\phi_2-\phi_1]
\label{eqn:int}
\end{equation}
For simplicity, we will now assume that $a_1$ and $a_2$ are equal and do not depend on $x$. We can then set
\begin{equation}
a_1 = a_2 = {\textstyle\frac1{\sqrt2}}
\end{equation}
by a suitable choice of units. Further writing
\begin{equation}
\begin{aligned}
\Delta l(x) & \equiv l_1(x)-l_2(x) \\
\Delta \phi & \equiv \phi_1-\phi_2
\end{aligned}
\end{equation}
the intensity becomes
\begin{equation}
I(x) = 1 + \cos[k\,\Delta l(x) - \Delta\phi]
\label{eqn:inten}
\end{equation}
This equation (eqn.~\ref{eqn:inten}) expresses the formation of an interference pattern on the screen as a result of the optical path difference $\Delta l(x)$ between the light rays.

For the interference pattern (eqn.~\ref{eqn:inten}) to be observable, it is essential that the extra phase $\Delta\phi$ is, if not constant, slowly varying. The individual phases $\phi_1$ and $\phi_2$ will in practice vary rapidly. It is because any light source, even a laser, is not perfectly monochromatic but has a nonzero frequency bandwidth $\Delta\nu$.  Frequency components differing by $\Delta\nu$ will go out of phase with each other over a time of order
\begin{equation}
\Delta\tau = \frac1{\Delta\nu} = \frac{\lambda ^2}{c \Delta \lambda}
\end{equation}
It is known as the coherence time. For lasers, the coherence time can be microseconds; for incoherent light, it is much shorter. The phase difference $\Delta\phi$ can still, however, be held constant by ensuring that $\phi_1=\phi_2$. In lasers, this condition follows from the process of stimulated emission that produces the light. Without a laser, $\phi_1=\phi_2$ can be realized by another technique, which is to take nearly monochromatic light (such as from a sodium lamp) and pass it through a narrow aperture before splitting it into two beams; this works if the aperture is small enough that the optical path length to the screen from different parts of the aperture differ by much less than a wavelength.

\subsection{Introducing incoherence:}
Now, suppose that a turbulent substance is introduced in the paths of the light beams, causing $\phi_1$ and $\phi_2$ to vary randomly. Then $\Delta\phi$ will also vary randomly. The interference pattern (eqn.~\ref{eqn:inten}) will fluctuate as $\Delta\phi$ varies, 
and as a consequence, average out to uniform. Remarkably, however, the pattern can still be reconstructed. Consider the product of intensities at two points on the screen, as follows.
\begin{equation}
\begin{aligned}
I(x) \, I(x') = 1
&+ \cos[k\,\Delta l(x)- \Delta \phi] + \cos[k\,\Delta l(x') - \Delta \phi] \\
&+ {\textstyle\frac12} \cos\left[ k\,\left(\Delta l(x)
+ \Delta l(x')\right) -  2\Delta \phi \right] \\
&+ {\textstyle\frac12} \cos\left[ k\,\left(\Delta l(x)
-\Delta l(x')\right) \right]
\end{aligned}
\end{equation}
This product will also fluctuate, but its time average will not be uniform because of the last cosine term, which has no $\Delta \phi$ dependence. The time average comes to
\begin{equation}
\left\langle I(x) \, I(x') \right\rangle = 1 +
{\textstyle\frac12} \cos\left[ k\,\left(\Delta l(x)
-\Delta l(x')\right) \right]
\label{eqn:corr}
\end{equation}
which is essentially the auto-correlation of the interference pattern for coherent light. Thus, the information on the interference of two beams generated by the beam splitter from a pseudo-thermal light source can still be retrieved by evaluating the second-order correlation. This result is called the \textbf{Hanbury Brown and Twiss (HBT) effect}.

The preceding can be generalized to extended sources. The amplitude (eqn.~\ref{eqn:ampl}) then generalizes to an integral of the form
\begin{equation}
A(x) = \int a_\mu \exp\left[i\phi_\mu - ik\,l_\mu(x)\right] \, d\mu
\label{eqn:extend}
\end{equation}
For many applications, $l_\mu(x)$ turns out to be linear in $x$, which conveniently makes $A(x)$ essentially a Fourier transform of $a_\mu$. The corresponding generalization of the intensity correlation (eqn.~\ref{eqn:corr}) then involves the absolute value of that Fourier transform.

In our experiment (see Fig.~\ref{fig:MI}), we used a lens to broaden a laser beam and also introduced a turbulent layer to produce an extended source with uncorrelated phases across it. The light from this extended source is then split into two beams. The next section discusses it in detail. 
\section{Experimental Setup}\label{sec:expt}
\begin{figure*}
	\centering
	\begin{subfigure}{0.50\linewidth}
		\includegraphics[width=\linewidth, height=2.7in]{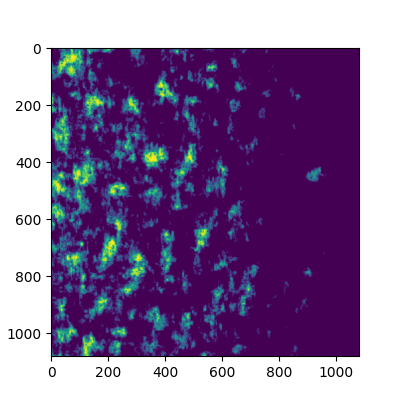}
	\end{subfigure}\hfill
	\begin{subfigure}{0.50\linewidth}
		\includegraphics[width=\linewidth, height=2.7in]{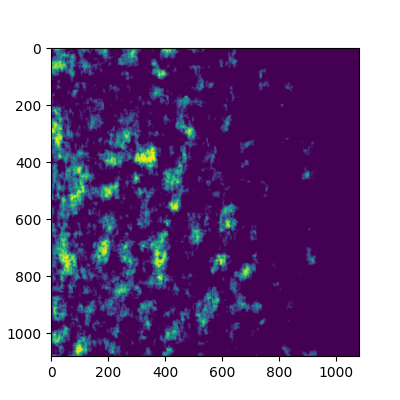}
	\end{subfigure}
	\begin{subfigure}{0.50\linewidth}
		\includegraphics[width=\linewidth, height=2.7in]{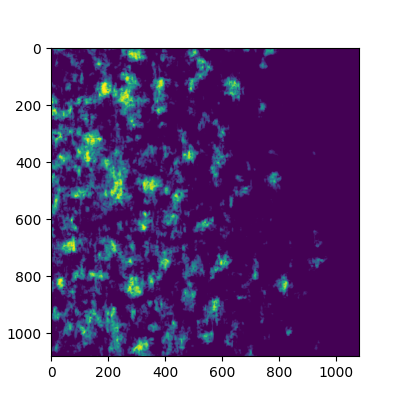}
	\end{subfigure}\hfill
	\begin{subfigure}{0.50\linewidth}
		\includegraphics[width=\linewidth, height=2.7in]{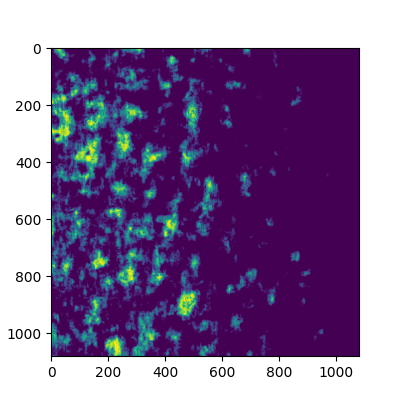}
	\end{subfigure}
	\begin{subfigure}{0.50\linewidth}
		\includegraphics[width=\linewidth, height=2.7in]{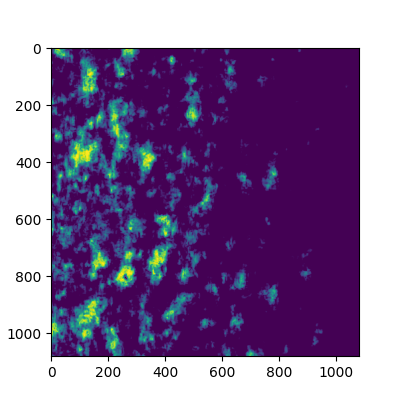}
	\end{subfigure}\hfill
	\begin{subfigure}{0.50\linewidth}
		\includegraphics[width=\linewidth, height=2.7in]{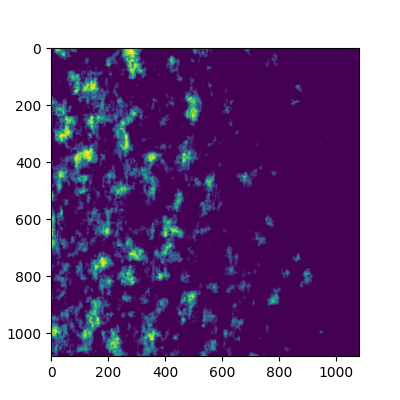}
	\end{subfigure}   
	\caption{Set of intensity distributions with pseudo thermal light. It is a two-channel distribution
		of pixels. So, it only has the distribution over pixels as a combination of green and blue colors.}
	\label{fig:fringes}   
\end{figure*}
\begin{figure}
	\centering
	\includegraphics[width=0.5\linewidth]{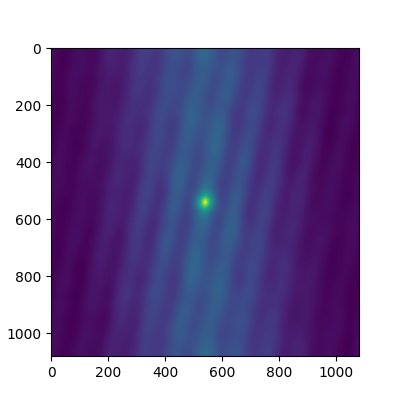}
	\caption{Average of auto-correlation of the Intensity distribution over many frames. Fig.~\ref{fig:fringes} is only one of them. This figure shows a clear pattern of interference with the pseudo-thermal light.}
	\label{fig:autocorre}
\end{figure}
For the work we report here, we have considered the basic design of the Michelson interferometer. The schematic organization of this experimental setup is depicted in Fig.~\ref{fig:MI}. Our experimental setup differs slightly from Hanbury Brown and Twiss' design, where they employed a mercury arc, a liquid filter, and a pinhole as the source, a beam-splitter, 
two photo-multipliers to detect photons, and a correlator [6]. In the work presented here, we have used a green laser with a wavelength of 532 nm. 
A lens is used before the beam splitter to disperse the light and produce an intensity distribution over a broad area of the screen. The beam-splitter (aligned at $45^0$ with respect to the laser beam direction) divides the light, causing it to reflect from two mirrors, resulting in interference fringes on the screen. We obtain vertical fringes by using tilted reflecting mirrors. The smartphone's CMOS camera is placed behind the screen to capture the intensity distribution. The purpose of the screen is to regulate and moderate the brightness of the intensity pattern so that the camera's detector does not get saturated and returns a clear intensity distribution.

Next, we convert coherent light from the laser into pseudo-thermal light. A square cuvette (of glass with a thickness of a few mm and a square cross-sectional dimension of 1 cm) containing distilled water and a small amount of milk is placed between the laser and the lens. Due to the existence of microscopic particles in the laser field, the phase of each wave varies randomly with time. In fig.~\ref{fig:MI}, it has been shown that when a beam of laser light passes through the cuvette, it disperses in a broader area due to the medium in the cuvette. The quantity of milk drop is controlled to ensure that the intensity distribution does not entirely vanish on-screen and that the intensity distribution on-screen changes over time (fluctuation in intensity distribution over screen). Fig.~\ref{fig:MII} depicts our typical laboratory setup for testing the HBT effect. Pseudo-thermal light can also be generated using a randomly moving or rotating thin, scratched, and transparent glass sheet in front of coherent light. In this case, random phase change would also occur over time and would result in fluctuating intensity distribution. 
\section{Result}\label{sec:result}
The CMOS camera records video of the random intensity distribution for some time interval, typically a few seconds. One of the benefits of using HBT correlation is that the recorded data can be stored on a hard disk for later use. So, the recorded video is split into multiple frames. Each video 
frame comprises three channels (red, green, and blue); however, only two 
are needed to represent the dark and bright intensity distribution. Fig.~\ref{fig:fringes} shows a few typical two-channel (green and blue) frames generated from the movie by time slicing. These frames represent the intensity distribution on the 1080$\times$1080 pixel grids of the CMOS camera produced by the pseudo-thermal light emanating from the 
intervening cuvette. As expected, these frames show no interference pattern. However, the time-averaged second-order correlation of these intensity distribution frames implemented through a sequence of numerical steps produces an interference pattern as suggested by eqn.~\ref{eqn:corr}. This numerical procedure, which basically calculates the time-averaged auto-correlation of these frames, can be enumerated as below: 
\begin{enumerate}
	\item[I.] Obtain the Fourier Transform of each frame obtained from the captured video.
	\item[II.] Compute the average of the square of these Fourier Transforms over all the frames
	\item[III.] Obtain the HBT correlation (eqn.~\ref{eqn:corr}) on the frames by computing 
	the inverse Fourier Transform of the average from step II.
\end{enumerate}
Fig.~\ref{fig:autocorre} depicts this pattern obtained by averaging over time-sliced photo frames generated from the video. This figure represents the HBT effect produced by the experimental setup (Fig.~\ref{fig:MII}).

In this article, we have demonstrated the HBT effect using the intensity correlation of a pseudo-thermal light with a variant of the Michelson Interferometer setup. It is worthwhile to note that, in astronomy, the HBT effect is used as II for the resolution of stellar objects with two-point 
correlation. In our future work, we plan to elaborate and expand this work wherein two pinholes (separated by a certain distance) acting as 
artificial ``laboratory stars" would be used to simulate a binary system. This system would not need a beam splitter. This experiment would attempt to validate our simulation results on the Spica system \cite{10.1093/mnras/stab2391}. 
The estimation (measurement) of the pinhole sizes and their separation using the intensity correlations would be tantamount to the extraction of source information of a stellar system using II. Similarly, we have also simulated the possibility of exoplanet detection on the wide binary system of Alpha Centauri AB \cite{10.1093/mnras/stac2433} with the proposition of masked apertures of the telescope objectives. So, a lab demonstration of the masked aperture can be explored.

\section{Conclusion}\label{sec:conclusion}
This study attempts to demonstrate the HBT effect in a college laboratory setting. The HBT effect shows that we can relax the strict requirement of coherence of the interfering beams as demanded by the Michelson Interferometer and yet be able to extract source information. It can be done by considering the averaged auto-correlation of intensity distribution produced by two secondary sources. The HBT effect is used in astronomy as Intensity Interferometry, where second-order (intensity) correlation is computed instead of first-order (amplitude) correlation. 
As a result, the smaller stellar objects are resolved with longer baselines. The flip side of this result, which is not easily achievable using Amplitude (Michelson) Interferometry, is that the phase information is lost. However, there are ways to extract the phase from the three-point HBT correlation using three secondary sources as well \cite{saha2020theory}. 

This article discusses and demonstrates several fundamental (both theoretical and experimental) aspects of Interference. Starting with coherent laser sources, it considers the general case of non-coherent sources. It discusses the theoretical background and practical procedure of extracting source information from the observed intensity distribution due to non-coherent sources. Thus, this article may serve as an exposition of fundamental Optics and cater to the universal goal of Quality Education (SDG-4) proposed by the United Nations.

\bigskip

\section*{Acknowledgement}
The authors of this article thank Akhileshwar Mishra and Navya Paul for their invaluable assistance in getting the instruments necessary for the experiments reported here. One of the authors (SS) acknowledges, hereby, the hospitality and support he received as a Visiting Associate at the Inter-University Centre for Astronomy and Astrophysics (IUCAA), Pune, during the preparation and submission of the first version of this manuscript. 

\bigskip
\bibliographystyle{apalike}
\bibliography{main}

\begin{thebibliography}{}

\bibitem[Abbott et~al., 2016]{abbott2016gw151226}
Abbott, B.~P., Abbott, R., Abbott, T., Abernathy, M., Acernese, F., Ackley, K.,
  Adams, C., Adams, T., Addesso, P., Adhikari, R., et~al. (2016).
\newblock Gw151226: observation of gravitational waves from a 22-solar-mass
  binary black hole coalescence.
\newblock {\em Physical review letters}, 116(24):241103.

\bibitem[Brown and Twiss, 1958]{brown1958interferometry}
Brown, R.~H. and Twiss, R. (1958).
\newblock Interferometry of the intensity fluctuations in light. ii. an
  experimental test of the theory for partially coherent light.
\newblock {\em Proceedings of the Royal Society of London. Series A.
  Mathematical and Physical Sciences}, 243(1234):291--319.

\bibitem[Brown and Twiss, 1954]{brown1954lxxiv}
Brown, R.~H. and Twiss, R.~Q. (1954).
\newblock Lxxiv. a new type of interferometer for use in radio astronomy.
\newblock {\em The London, Edinburgh, and Dublin Philosophical Magazine and
  Journal of Science}, 45(366):663--682.

\bibitem[Brown and Twiss, 1956]{brown1956correlation}
Brown, R.~H. and Twiss, R.~Q. (1956).
\newblock Correlation between photons in two coherent beams of light.
\newblock {\em Nature}, 177(4497):27--29.

\bibitem[Brown et~al., 1957]{brown1957interferometry}
Brown, R.~H., Twiss, R.~Q., and surName, g. (1957).
\newblock Interferometry of the intensity fluctuations in light-i. basic
  theory: the correlation between photons in coherent beams of radiation.
\newblock {\em Proceedings of the Royal Society of London. Series A.
  Mathematical and Physical Sciences}, 242(1230):300--324.

\bibitem[Davis et~al., 1999]{davis1999sydney}
Davis, J., Tango, W., Booth, A., Brummelaar, T.~t., Minard, R., and Owens, S.
  (1999).
\newblock The sydney university stellar interferometer—i. the instrument.
\newblock {\em Monthly Notices of the Royal Astronomical Society},
  303(4):773--782.

\bibitem[Glauber, 1963]{glauber1963quantum}
Glauber, R.~J. (1963).
\newblock The quantum theory of optical coherence.
\newblock {\em Physical Review}, 130(6):2529.

\bibitem[Hanbury~Brown et~al., 1974]{hanbury1974angular}
Hanbury~Brown, R., Davis, J., and Allen, L. (1974).
\newblock The angular diameters of 32 stars.
\newblock {\em Monthly Notices of the Royal Astronomical Society},
  167(1):121--136.

\bibitem[Hanbury~Brown et~al., 1952]{hanbury1952apparent}
Hanbury~Brown, R., Jennison, R., and Gupta, M.~D. (1952).
\newblock Apparent angular sizes of discrete radio sources: Observations at
  jodrell bank, manchester.
\newblock {\em Nature}, 170(4338):1061--1063.

\bibitem[Hanbury~Brown and Twiss, 1979]{hanbury1979test}
Hanbury~Brown, R. and Twiss, R.~Q. (1979).
\newblock A test of a new type of stellar interferometer on sirius.
\newblock In {\em A Source Book in Astronomy and Astrophysics, 1900--1975},
  pages 8--12. Harvard University Press.

\bibitem[Michelson and Pease, 1921]{michelson1921measurement}
Michelson, A.~A. and Pease, F.~G. (1921).
\newblock Measurement of the diameter of alpha-orionis by the interferometer.
\newblock {\em Proceedings of the National Academy of Sciences}, 7(5):143--146.

\bibitem[Rai et~al., 2021]{10.1093/mnras/stab2391}
Rai, K.~N., Basak, S., and Saha, P. (2021).
\newblock {Radius measurement in binary stars: simulations of intensity
  interferometry}.
\newblock {\em Monthly Notices of the Royal Astronomical Society},
  507(2):2813--2824.

\bibitem[Rai et~al., 2022]{10.1093/mnras/stac2433}
Rai, K.~N., Sarangi, S., Saha, P., and Basak, S. (2022).
\newblock {Simulations of astrometric planet detection in Alpha Centauri by
  intensity interferometry}.
\newblock {\em Monthly Notices of the Royal Astronomical Society},
  516(2):2864--2875.

\bibitem[Saha, 2020]{saha2020theory}
Saha, P. (2020).
\newblock The theory of intensity interferometry revisited.
\newblock {\em arXiv preprint arXiv:2009.07284}.

\end{thebibliography}
\end{document}